%% file: phichic_prd.tex
\RequirePackage{lineno}
\documentclass[aps,prd,twocolumn,showpacs,byrevtex]{revtex4}
\usepackage[T1]{fontenc}
\usepackage{times}
\usepackage{amsmath}
\usepackage{amssymb}
\usepackage{graphicx}
\usepackage{color}
\usepackage{multirow}
\usepackage{setspace}
\usepackage{lineno}
\usepackage{verbatim}
\newcommand{\pp}{\pi^+\pi^-}
\newcommand{\kk}{K^+ K^-}
\newcommand{\LL}{\ell^+ \ell^-}
\newcommand{\ee}{e^+e^-}

\newcommand{\uu}{\mu^+\mu^-}
\newcommand{\GG}{\gamma\gamma}
\newcommand{\jpsi}{J/\psi}
\newcommand{\lamlam}{\Lambda_c^+\bar{\Lambda}_c^-}
\newcommand{\incfig}[2]{\includegraphics[width=#1\textwidth]{#2}}

\begin{document}
\graphicspath{{figure/}}
\DeclareGraphicsExtensions{.eps,.png,.ps}
\title{\boldmath
Observation of $\ee\to \phi\chi_{c1}$ and $\phi\chi_{c2}$ at
$\sqrt{s}=4.600$~GeV} \vspace{0.4cm}
\input{authors_feb2017}

\begin{abstract}

Using a data sample collected with the BESIII detector operating
at the BEPCII storage ring at a center-of-mass energy of
$\sqrt{s}=4.600$~GeV, we search for the production of $\ee\to
\phi\chi_{c0,1,2}$. A search is also performed for the charmonium-like
state $X(4140)$ in the radiative transition $\ee\to \gamma X(4140)$ with $X(4140)$
subsequently decaying into $\phi\jpsi$. The processes $\ee\to
\phi\chi_{c1}$ and $\phi\chi_{c2}$ are observed for the first
time, each with a statistical significance of more than
$10\sigma$, and the Born cross sections are measured to be
$(4.2^{+1.7}_{-1.0}\pm 0.3)$~pb and $(6.7^{+3.4}_{-1.7}\pm
0.5)$~pb, respectively, where the first uncertainties are
statistical and the second systematic. No significant signals are
observed for $\ee\to \phi\chi_{c0}$ and $\ee\to \gamma X(4140)$
and upper limits on the Born cross sections at $90\%$ confidence
level are provided at $\sqrt{s}=4.600$~GeV.

\end{abstract}

\pacs{14.40.Pq, 14.40.Rt, 13.66.Bc}
\maketitle
\section{INTRODUCTION}

In recent years, many charmonium-like states have been observed
experimentally, whose characters are different from the
predictions of the charmonium states in the potential model. The
$X(3872)$ was first observed by the Belle Collaboration in
$B^{\pm}\to K^{\pm}\pp\jpsi$~\cite{3872_1} and was subsequently
confirmed by several other
experiments~\cite{3872_2,3872_3,3872_4,3872_5}. The vector states
$X(4260)$, $X(4360)$, and $X(4660)$, sometimes called the $Y(4260)$, $Y(4360)$,
and $Y(4660)$, were discovered by the {\em BABAR}, Belle, and CLEO Collaborations
via their decays into low-mass charmonium states $\pp\jpsi$ or
$\pp\psi(3686)$~\cite{4260_1,4260_2,4260_3,4360,4660}. Some
charged charmonium-like states and their neutral partners, such as
$Z_c(3900)$, $Z_c(3885)$, $Z_c(4020)$, $Z_c(4025)$, $Z_c(4200)$
have been also observed by several
experiments~\cite{3900_1,3900_2,3900_3,3900_4,3885_1,3885_2,4020_1,4020_2,4025_1,4025_2,4200_1}.
There are many theoretical interpretations on the nature of these
$XYZ$ states, such as molecular, hybrid, or multi-quark states,
threshold enhancements, or some other
configurations~\cite{configuration}. However, the nature of these
states is still unclear. Due to the richness of $XYZ$ states above
the open charm threshold, searching for new decay modes of these
states and measuring their line shape precisely will provide
helpful information to determine the properties of them.

The authors of Ref.~\cite{predict} predicted a sizable coupling
between the $X(4260)$ and the $\omega\chi_{c0}$ channel by
considering the threshold effect of the $\omega\chi_{c0}$. The
BESIII Collaboration measured the cross sections of $\ee\to
\omega\chi_{c0,1,2}$ at center-of-mass (c.m.) energies between
$4.23$ and $4.60$~GeV and determined the mass of an intermediate
resonance to be about $4226$~MeV/$c^2$, assuming that the
$\omega\chi_{c0}$ signals come from a single
resonance~\cite{c0,c12}. These resonant parameters are also
inconsistent with those obtained by fitting a single resonance to
the $\pp\jpsi$ cross section~\cite{4260_1,4260_2}. Recently,
BESIII Collaboration precisely measured the cross section of
$\ee\to \pp\jpsi$ in the relevant mass range and observed two
resonant structures whose masses are determined to be $4224$ and
$4319$~MeV/$c^2$~\cite{ppjpsi}. The mass of the first state is
lower than that from {\em BABAR} and Belle measurements
corresponding to the $X(4260)$. The fact that the parameters of
the $X(4260)$ agree with the structure observed by BESIII
Collaboration in $\ee\to \omega\chi_{c0}$ suggests that the
$X(4260)$ have multiple decay modes. Considering that $\omega$ and
$\phi$ mesons have the same spin, parity, and isospin,
$\omega\chi_{cJ}$ and $\phi\chi_{cJ}$ may have a similar
production mechanism. Therefore, we study and measure the cross
sections of $\ee\to \phi\chi_{c0,1,2}$.

The $X(4140)$, sometimes called the $Y(4140)$, was
first reported by the CDF experiment in the
decay $B^{+}\to \phi\jpsi K^{+}$~\cite{4140_CDF}. However, the
existence of the $X(4140)$ was neither confirmed by the
Belle~\cite{4140_Be} and {\em BABAR}~\cite{4140_BA} Collaborations
in the same process, nor by Belle Collaboration in two-photon
production~\cite{4140_Be}. Recently, the CMS~\cite{4140_CMS} and
{D{\O}~\cite{4140_D0} Collaborations reported the observation of
the $X(4140)$ with resonant parameters being consistent with those
of the CDF measurement. More recently, the LHCb Collaboration
observed the $X(4140)$ with a statistical significance of
$8.4\sigma$ using a data sample of $3~{\rm fb}^{-1}$ $pp$
collision in the same process~\cite{4140_LHCb}, using a full
amplitude analysis. BESIII Collaboration has searched for the
$X(4140)$ in the process $\ee\to \gamma\phi\jpsi$ with data
samples at c.m.\ energies $\sqrt{s}=4.23$, $4.26$, and
$4.36$~GeV~\cite{4140_BESIII}, but no obvious signal has been
observed.

In this article, we present the results of a study of $\ee\to
\phi\chi_{c0,1,2}$ and a search for the $X(4140)$ in process
$\ee\to \gamma X(4140)\to \gamma\phi\jpsi$, based on an $\ee$
annihilation data sample collected with the BESIII
detector~\cite{bes3} at $\sqrt{s}=4.600$~GeV. The c.m.\ energy of
the data sample is determined with a precision of
$0.8$~MeV~\cite{cmsenergy} using di-muon events. The integrated
luminosity of the sample is measured using large-angle Bhabha
scattering to be $567$~$\rm pb^{-1}$ with a precision of
1.0\%~\cite{luminosity}.

\section{DETECTOR AND MONTE CARLO SAMPLES}

The Beijing Spectrometer III (BESIII) detector, described in
detail in Ref.~\cite{bes3}, is a magnetic spectrometer operating
at the Beijing Electron-Positron collider (BEPCII), which is a
double-ring $\ee$ collider with a c.m.\ energy range from $2.0$ to
$4.6$~GeV. The cylindrical core of the BESIII detector consists of
a helium-based main drift chamber (MDC), a plastic scintillator
time-of-flight (TOF) system, and a CsI(Tl) electromagnetic
calorimeter (EMC) that are all enclosed in a superconducting
solenoid magnet providing a $1.0$~T magnetic field. The magnet is
supported by an octagonal flux-return yoke with modules of
resistive plate muon counters (MUC) interleaved with steel. The
acceptance of the MDC for charged tracks is $93\%$ of $4\pi$ solid
angle. It provides a charged particle momentum resolution of
$0.5\%$ at $1.0$~GeV/$c$ and ionization energy loss ($dE/dx$)
measurements with resolution better than $6\%$. The time
resolution of the TOF is $80\;(110)$~ps for the barrel (end caps)
and the EMC measures photon energy with a resolution of
$2.5\%\;(5\%)$ at $1.0$~GeV in the barrel (end caps). The MUC
provides a position resolution of $2$~cm and detects muon tracks
with momenta higher than $0.5$~GeV/$c$.

The optimization of event selection, determination of the
detection efficiency and estimation of the backgrounds are
performed using the {\sc geant4}-based~\cite{geant4} Monte Carlo
(MC) simulation software {\sc boost}~\cite{boost}. It includes the
geometric and material description for the BESIII detector and a
simulation of the detector response. Signal MC samples of $\ee\to
\phi\chi_{c0,1,2}$ and $\ee\to \gamma X(4140)\to \gamma\phi\jpsi$
are generated at $\sqrt{s}=4.600$~GeV, where each sample contains
$10^{5}$ events. Both $\chi_{c1}$ and $\chi_{c2}$ states are
reconstructed via $\chi_{c1,2}\to \gamma\jpsi$, $\jpsi\to
\LL~(\ell=e$ or $\mu)$, and $\phi$ via its decay to $\kk$. For $\ee\to \phi\chi_{c0}$,
since the branching fraction of $\chi_{c0}\to \gamma\jpsi$, with
$\jpsi\to \LL$ is smaller than those of $\chi_{c0}\to \pp$, $\kk$,
$\pp\pp$, and $\kk\pp$, the $\chi_{c0}$ state is reconstructed with
the latter four channels. Initial state radiation effects are simulated with {\sc
kkmc}~\cite{kkmc}, where the production cross sections are assumed
to follow the line shape of the $X(4660)$~\cite{4660}, modified by
a phase space factor. Final state radiation effects associated
with charged particles are handled with {\sc
photos}~\cite{photos}.

An ``inclusive'' MC sample is also generated with an integrated
luminosity equivalent to that of the data sample. QED events,
$\ee\to \ee$, $\uu$, and $\GG$, are generated with {\sc
babayaga}~\cite{babayaga}. The processes including an intermediate
$D_{(s)}^{(*)}$ meson such as $\ee\to D\bar{D}$,
$D^{*}\bar{D^{*}}$, $D\bar{D^{*}}+c.c.$, $D_{s}^+D_{s}^-$,
$D_{s}^+D_{s}^{*-}+c.c.$, and $D_{s}^{*+}D_{s}^{*-}$, the known
charmonium production processes, and the process $\ee\to \lamlam$
with all their known decays are generated using {\sc
evtgen}~\cite{evtgen}. The unmeasured but possible decays
associated to charmonium states are generated with {\sc
lundcharm}~\cite{lundcharm} and other hadronic events are
generated with {\sc pythia}~\cite{pythia}.


\section{\boldmath $\ee\to \phi\chi_{c1}$ and $\phi\chi_{c2}$}

\subsection{Event Selection}

The final states for $\ee\to \phi\chi_{c1}$ and $\phi\chi_{c2}$
are $\gamma \kk\LL$. For each charged track in the MDC, the polar
angle must satisfy $|\cos\theta|<0.93$ and the point of closest
approach to the $\ee$ interaction point must be within $\pm 10$~cm
in the beam direction and within $1$~cm in the plane perpendicular
to the beam direction. We require that there are at least three
candidate charged tracks in the final state. Leptons from $\jpsi$
decays can be separated from other tracks kinematically, hence the
two tracks with momenta greater than $1.0$~GeV/$c$ and opposite
charge are assumed to be leptons. The energy deposited in the EMC
is used to separate electrons from muons. For muon candidates, the
deposited energy is required to be less than $0.6$~GeV, while for
electron candidates it is required to be greater than $1.0$~GeV.
The momenta of the kaons are about $0.2$~GeV/$c$ in the laboratory
frame, and low momentum kaons affect the reconstruction efficiency
significantly. To increase the efficiency, only one kaon is
required to be reconstructed and pass the particle identification
(PID) requirements. For each charged track with low momentum, the
PID probability $Prob_{i}(i=\pi,K)$ of each particle hypothesis is calculated,
combining the $dE/dx$ and TOF information. Here we require $Prob_{K}>Prob_{\pi}$.

Photon candidates are reconstructed from showers in the EMC
crystals. Each photon is required to have an energy deposition
above $25$~MeV in the barrel of the EMC ($|\cos\theta|<0.80$) or
$50$~MeV in the end caps ($0.86<|\cos\theta|<0.92$). To exclude
showers due to bremsstrahlung radiation from charged tracks, the
angle between the shower position and the nearest charged tracks,
extrapolated to the EMC, must be greater than $20$ degrees. The
timing information from the EMC is restricted to be $0 \le t \le
700$~ns to suppress electronic noise and energy deposits unrelated
to the event. At least one photon candidate is required.

In order to improve the mass resolution and suppress backgrounds,
a one-constraint ($1$C) kinematic fit is performed under the
$\ee\to \gamma K^{\pm}K^{\mp}_\text{miss}\LL$ hypothesis by
constraining the mass of the missing track to be the kaon mass. If
there are two kaons or more than one candidate photon, the
combination of $\gamma K^{\pm}K^{\mp}_\text{miss}\LL$ with the
least $\chi^2$ is accepted. The $\chi^2$ of the kinematic fit is
required to be less than $20$.

With all of the above selection criteria being applied, the 
invariant mass distribution of $M(\kk)$ versus $M(\LL)$
and the corresponding $1$-D projections for data are shown in
Fig.~\ref{fig:scatter}(a-c). By default, $M$ denotes the invariant 
mass. Obvious signals can be seen in the
$\phi$ and $\jpsi$ mass windows, which are defined as $0.995\leq
M(\kk)\leq 1.048$~GeV/$c^2$ and $3.046\leq M(\LL)\leq
3.150$~GeV/$c^2$, respectively. The mass windows of the $\phi$ and
$\jpsi$ are four times the full width at half maximum of the
invariant mass distributions of signal events from the MC
simulation. The distribution of $M(\kk)$ versus
$M(\gamma\jpsi)$ after the $\jpsi$ mass window requirement is shown in
Fig.~\ref{fig:scatter}(d). The signal regions of $\chi_{c1}$ and
$\chi_{c2}$ states are set to be $[3.49, 3.53]$ and $[3.54,
3.58]$~GeV/$c^2$, respectively. Significant accumulations of
events can be seen in the intersections of the signal regions.

\begin{figure*}[htbp]
\incfig{0.42}{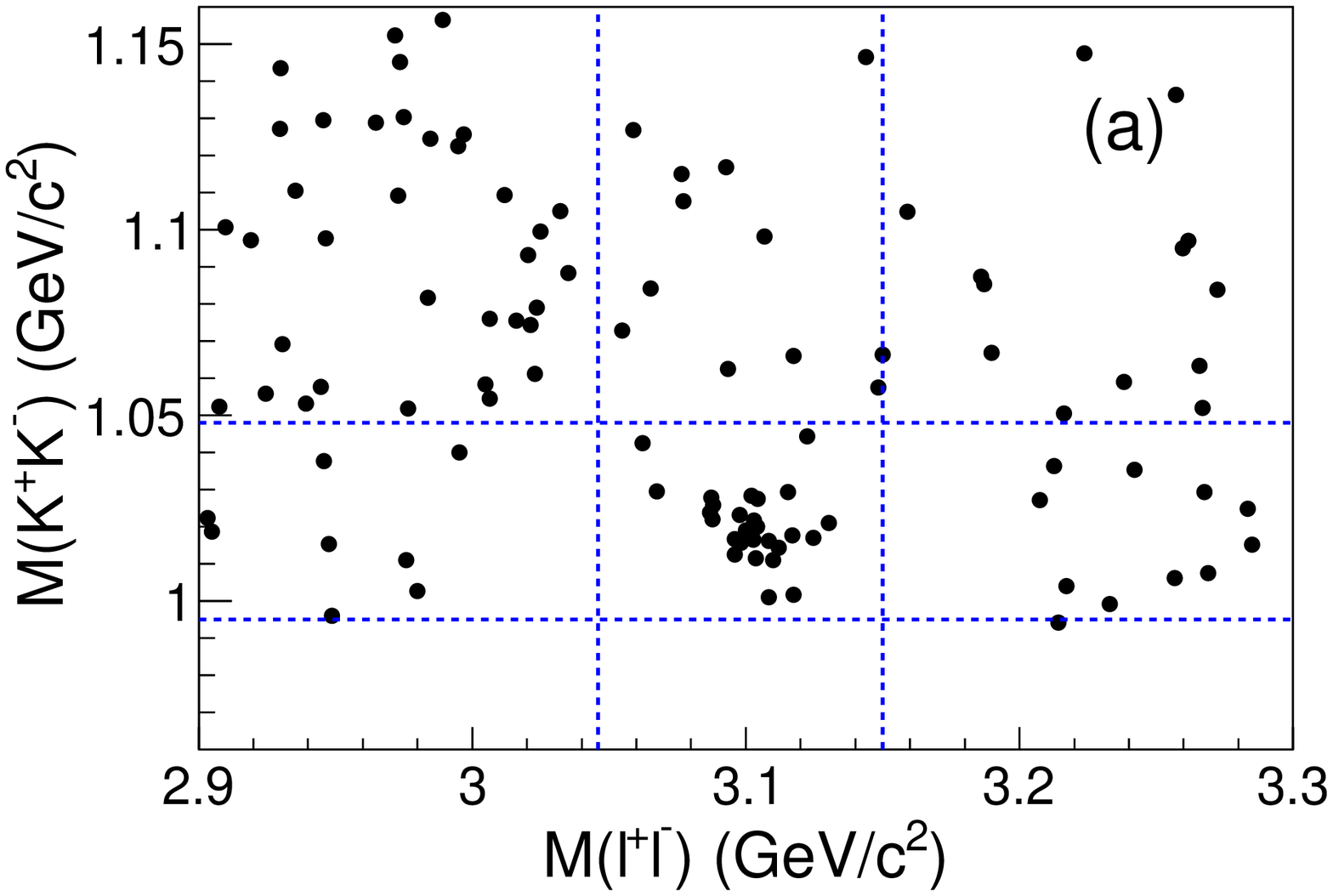}
\incfig{0.42}{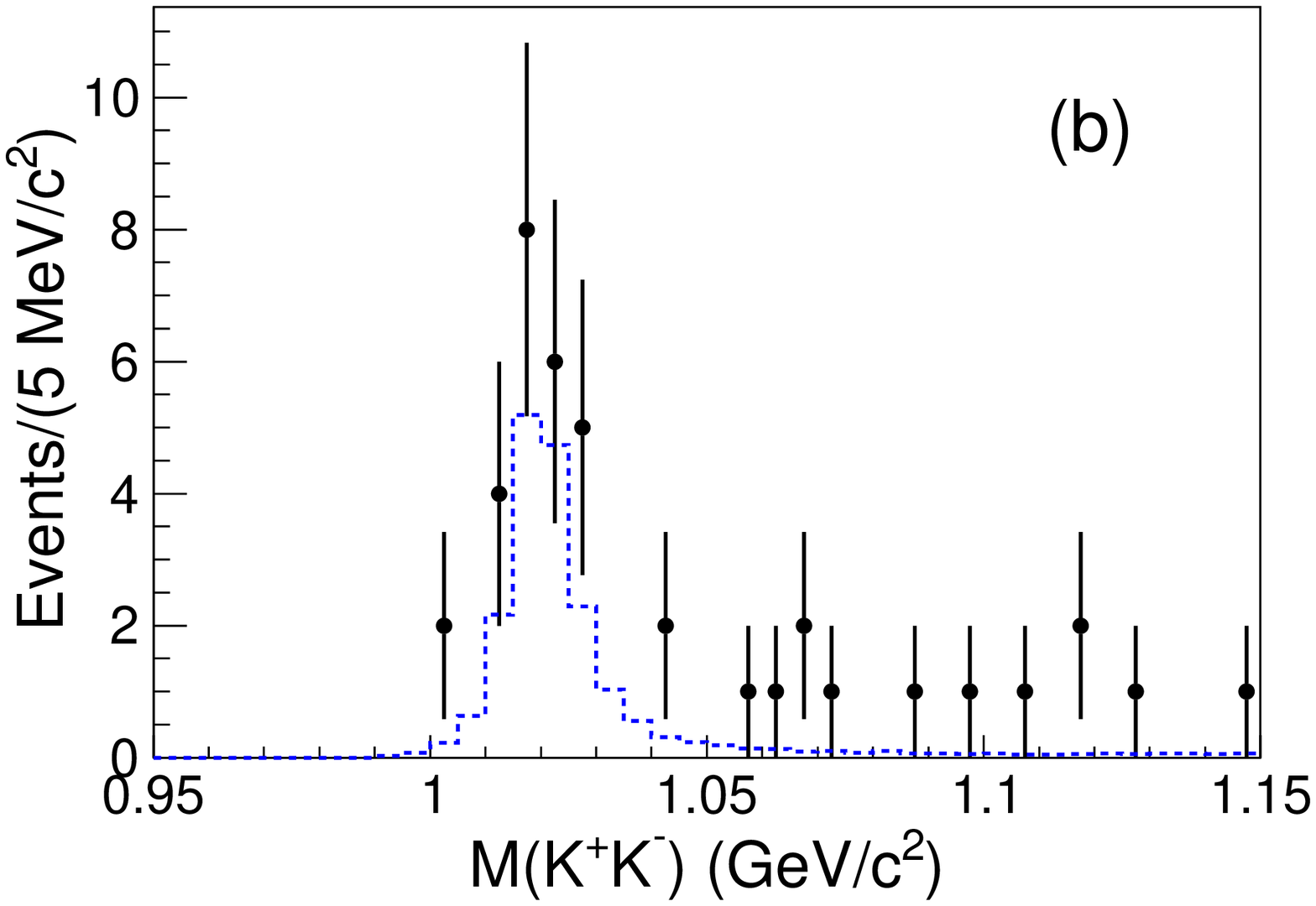}\\
\incfig{0.42}{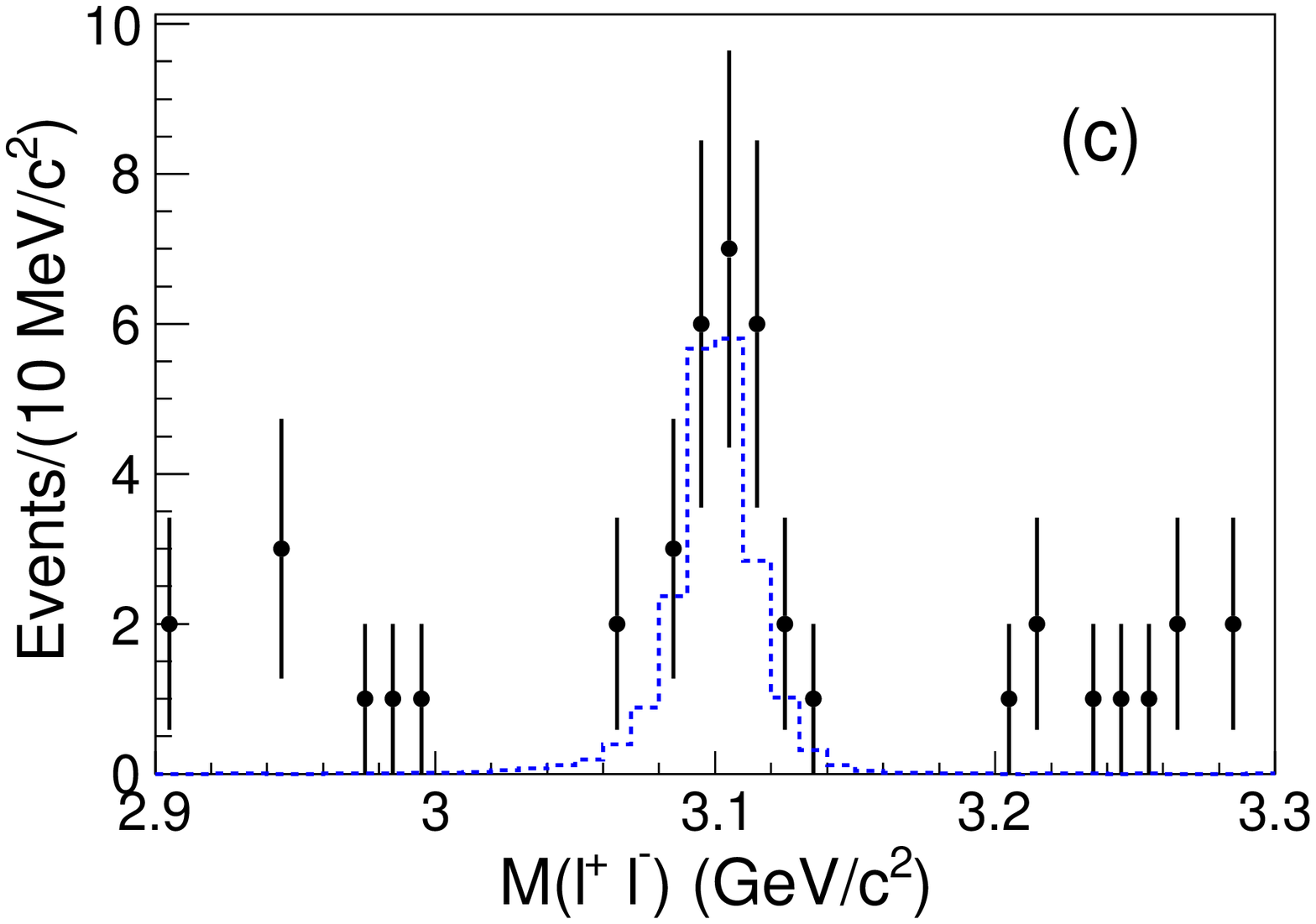}
\incfig{0.42}{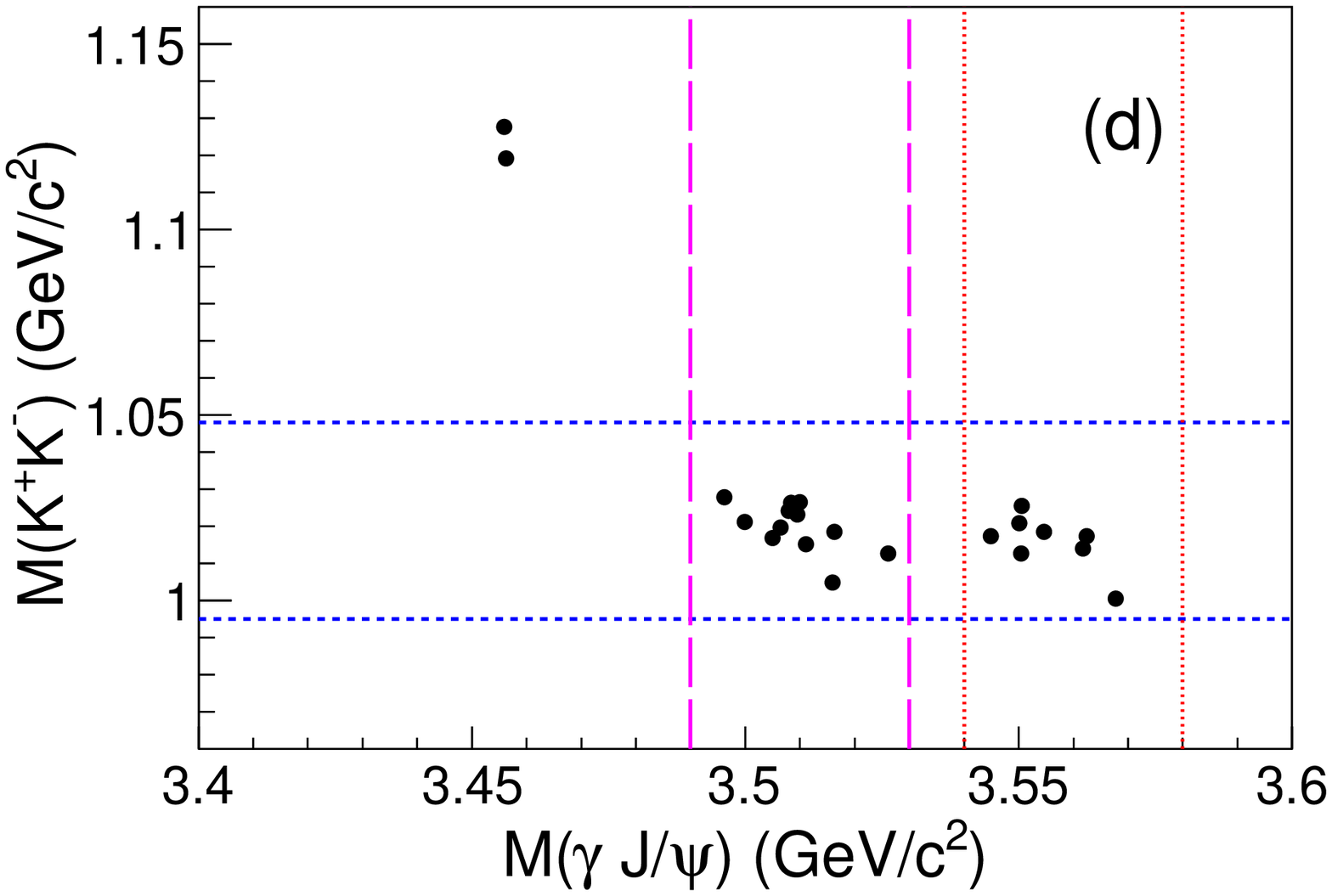}\\
\caption{(Color online) (a) Distribution of $M(\kk)$ versus
$M(\LL)$, (b) the projection along $M(\kk)$ in the $\jpsi$ mass
window, (c) the projection along $M(\LL)$ in the $\phi$ mass
window, and (d) distribution of $M(\kk)$ versus $M(\gamma\jpsi)$
in the $\jpsi$ mass window for data at $\sqrt{s}=4.600$~GeV. The
blue dashed lines represent the mass windows of the $\phi$ and
$\jpsi$ in plot (a) and (d). The blue dashed histograms in (b) and
(c) represent the MC simulated shapes of $M(\kk)$ and $M(\LL)$,
respectively, which have been normalized to the measured Born
cross sections. The magenta long-dashed and red dotted lines in
(d) represent the signal regions of the $\chi_{c1}$ and
$\chi_{c2}$, respectively.} \label{fig:scatter}
\end{figure*}

The same selection criteria are applied to the inclusive MC sample
to investigate possible background contributions. No events meet
the requirements. Furthermore, exclusive MC samples for several
processes, such as $\ee\to \kk\jpsi$, $\phi\pp$, $\kk\pp$,
$\kk\kk$, and $\kk\pp\pi^0$, which are potential background
channels but not included in the inclusive MC samples, are
generated separately. Each sample contains more than one million
events (corresponding to a cross section of $2$~nb at the current
luminosity). The cross sections of these processes have been
measured to be on the order of a few or a few tens of
pb~\cite{bkg1,bkg2,bkg3,bkg4} in the energy range of interest. We
find that the dominating background events originate from $\ee\to
\kk\jpsi$ in combination with a photon from initial state
radiation. Using the cross section of $\ee\to \kk\jpsi$ at
$\sqrt{s}=4.600$~GeV measured by BESIII~\cite{bkg1}, the numbers
of background events for the $\chi_{c1}$ and $\chi_{c2}$ channels
normalized to the luminosity of the data sample are estimated to
be $0.014$ and $0.002$, respectively. Simulation studies for all
possible backgrounds show that less than $0.2\%$ of the total
candidate events are from background contributions.

\subsection{Cross Sections}

The distribution of $M(\gamma\jpsi)$ after all event
selection requirements is shown in Fig.~\ref{fig:mgjpsi}. The
$\chi_{c1}$ and $\chi_{c2}$ signal regions are defined as $[3.49,
3.53]$ and $[3.54, 3.58]$~GeV/$c^{2}$, respectively. $12$ and $8$
events, respectively, are observed by counting the number of
events located in the $\chi_{c1}$ and $\chi_{c2}$ signal regions.

Assuming that the number of signal and background events both
follow a Poisson distribution, the confidence interval
$[\mu_{a},\mu_{b}]$ with confidence level $\gamma = 0.6827$,
should satisfy the formulas
\begin{equation}
\label{f1}
\int_{\mu=0}^{\mu_{a}}\sum\limits_{n=0}^{N}P(n,\mu)\cdot P((N-n),b)d\mu=\frac{1-\gamma}{2}=0.1587,
\end{equation}
\begin{equation}
\label{f2}
\int_{\mu=0}^{\mu_{b}}\sum\limits_{n=0}^{N}P(n,\mu)\cdot P((N-n),b)d\mu=\frac{1+\gamma}{2}=0.8413,
\end{equation}
where $P(n,\mu)=\frac{1}{n!}\mu^{n}e^{-\mu}$ is the probability
density function of a Poisson distribution, $N$ is the number of
the events observed in the signal region, $n$ is the number of the
signal events, $\mu$ is the expected number of signal events, $b$
is the expected number of background events, which is estimated
using the dedicated background MC samples. The signal yields of
the $\chi_{c1}$ and $\chi_{c2}$ channels are obtained to be
$12.0^{+4.6}_{-2.6}$ and $8.0^{+4.0}_{-2.0}$, respectively. The
p-value can be obtained by calculating the probability of the
expected number of background events to fluctuate to the number of
observed events or more in the signal regions assuming a Poisson
distribution. The p-value is $1.17\times10^{-31}$ for $\chi_{c1}$
and $6.34\times10^{-27}$ for $\chi_{c2}$, corresponding to
statistical significances of $11.6\sigma$ and $10.6\sigma$,
respectively.

\begin{figure}[htbp]
\incfig{0.42}{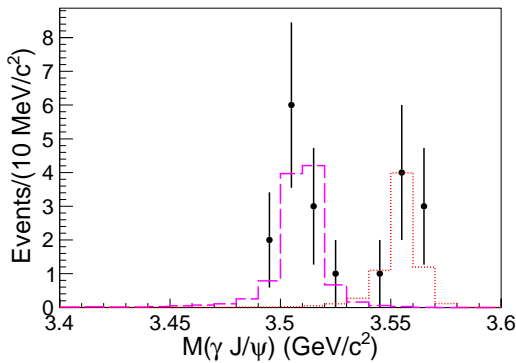}\\
\caption{(Color online) Distribution of $M(\gamma\jpsi)$,
after all requirements, for data at $\sqrt{s}=4.600$~GeV.
The markers with error bars are for data.
The magenta long-dashed and red dotted histograms are the shapes
of the $\chi_{c1}$ and $\chi_{c2}$ signals from MC simulation,
respectively, normalized to the measured Born cross sections.}
\label{fig:mgjpsi}
\end{figure}

The Born cross sections are calculated according to
\begin{equation}
\label{equ:born} \sigma^{\rm B}=\frac{N^{\rm sig}}{\mathcal{L}_{\rm int}~(\epsilon_e\mathcal{B}_e+\epsilon_{\mu}\mathcal{B}_{\mu})\mathcal{B}_{\chi_{c}}(1+\delta)(1+\delta^{\rm vac})},
\end{equation}
where $N^{\rm sig}$ is the number of the signal events,
$\mathcal{L}_{\rm int}$ is the integrated luminosity, $\epsilon_e$
and $\epsilon_\mu$ are the selection efficiencies for the $\ee$
and $\uu$ modes, respectively, and are listed in
Table~\ref{tab:cross}, $\mathcal{B}_e$ is the branching fraction
$\mathcal{B}(\jpsi\to \ee)$, $\mathcal{B}_\mu$ is the branching
fraction $\mathcal{B}(\jpsi\to \uu)$, $\mathcal{B}_{\chi_{c}}$ is
the branching fraction $\mathcal{B}(\chi_{c1,2}\to
\gamma\jpsi)\mathcal{B}(\phi\to \kk)$, $(1+\delta)$ is the
radiative correction factor, and $(1+\delta^{\rm vac})$ is the
vacuum polarization factor. We assume that the cross section for
$\ee\to \phi\chi_{c1,2}$ follows the $X(4660)$ line
shape~\cite{4660} modified by a two-body phase space factor,
\begin{equation}
\label{BW}
BW(\sqrt{s})=\frac{\Gamma_{ee}\mathcal{B}(\phi\chi_{c1,2})\Gamma}{(s-M^2)^2+(M\Gamma)^2}\cdot\frac{\Phi(\sqrt{s})}{\Phi(M)},
\end{equation}
where $BW$ is a Breit-Wigner function, the mass ($M$) and width
($\Gamma$) are taken from the Particle Data Group~\cite{pdg},
$\Gamma_{ee}$ is the partial width to $\ee$,
$\mathcal{B}(\phi\chi_{c1,2})$ is the branching fraction of
$X(4660)\to \phi\chi_{c1,2}$, and
$\Phi(\sqrt{s})=\frac{q}{\sqrt{s}}$ is the phase space factor for
an $S$-wave two-body system, where $q$ is the $\phi$ momentum in
the $\ee$ c.m.\ frame (with $\hbar=c=1$). The radiative correction
factor is obtained with a QED calculation~\cite{radiators}, using
the Breit-Wigner parameters of $X(4660)$~\cite{4660} as input. The
vacuum polarization factor $(1+\delta^{\rm vac})=1.055$ is taken
from Ref.~\cite{vacuum} and its uncertainty is negligible compared
with other uncertainties.

The Born cross sections of $\ee\to \phi\chi_{c1}$ and
$\phi\chi_{c2}$ at $\sqrt{s}=4.600$~GeV are measured to be
$4.2^{+1.7}_{-1.0}$ and $6.7^{+3.4}_{-1.7}$~pb, respectively. The
numbers used in the calculation and the results are listed in
Table~\ref{tab:cross}.

\begin{table}[hbp]
\caption{\small{The efficiencies ($\epsilon_{e}$ and
$\epsilon_{\mu}$), the radiative correction factor $(1+\delta)$,
the number of signal events $(N^{\rm sig})$, the Born cross
section $(\sigma^{\rm B})$, and the statistical significance for
$\ee\to \phi\chi_{c1}$ and $\phi\chi_{c2}$.}} \label{tab:cross}
\begin{tabular}{c c c c c c}
  \hline
  \hline
  Channel & $\epsilon_{e}(\epsilon_{\mu})(\%)$ & $1+\delta$ & $\rm N^{\rm sig}$ & $\rm \sigma^{\rm B}(\rm pb)$ & Significance\\
  \hline
  $\phi\chi_{c1}$   &  $28.5(38.6)$  &  $0.73$  &  $12.0^{+4.6}_{-2.6}$    & $4.2^{+1.7}_{-1.0}$  & $11.6\sigma$ \\
  $\phi\chi_{c2}$   &  $21.7(29.6)$  &  $0.71$  &  $8.0^{+4.0}_{-2.0}$     & $6.7^{+3.4}_{-1.7}$  & $10.6\sigma$ \\
  \hline
  \hline
\end{tabular}
\end{table}

\section{\boldmath $\ee\to \phi\chi_{c0}$}

\subsection{Event Selection}

\subsubsection{$\chi_{c0}\to \pp/\kk$}

For the decay modes $\chi_{c0}\to \pp/\kk$, we require that there
are three charged particle tracks for which the selection criteria
are the same as described above for the $\phi\chi_{c1}$ and
$\phi\chi_{c2}$ analyses. Similarly, we require only one kaon from
$\phi$ decays to be reconstructed and pass the PID requirement.
The tracks from $\chi_{c0}$ decays can be kinematically separated
from kaons from $\phi$ decays, hence the two oppositely charged
tracks with momenta greater than $1.0$~GeV/$c$ are assumed to be
$\pp$ or $\kk$ pairs from $\chi_{c0}$ decays. To separate
$\chi_{c0}\to \kk$ from $\chi_{c0}\to \pp$, a $1$C kinematic fit
is performed with the $\ee\to K^{\pm}K^{\mp}_\text{miss}\pp$ or
$K^{\pm}K^{\mp}_\text{miss}\kk$ hypothesis by constraining the
mass of the missing track to the kaon mass. If
$\chi^2(\chi_{c0}\to \pp)<\chi^2(\chi_{c0}\to \kk)$, the event is
identified as originating from $\chi_{c0}\to \pp$, otherwise from
$\chi_{c0}\to \kk$. The $\chi^2$ of the kinematic fit is required
to be less than $20$. If there is more than one kaon from the
$\phi$ decay identified, the combination with the least $\chi^2$
is retained.

To select signal events, we define the $\phi$ mass window as four
times the full width at half maximum of the distribution of
$M(\kk)$ of signal events from the MC simulation, resulting in
the requirement that $1.001\leq M(\kk)\leq 1.038$~GeV/$c^2$.
Figure~\ref{fig:scatter_kk/pp} shows the distributions of $M(\kk)$
for low momentum tracks versus $M(\pp/\kk)$ for high momentum
tracks from the data sample, as well as the $1$-D projections. No
obvious $\chi_{c0}$ signals are observed. By studying the
inclusive MC sample, we find that more than $90\%$ of background
events are from $\ee\to \phi\kk$.

\begin{figure*}[htbp]
\incfig{0.42}{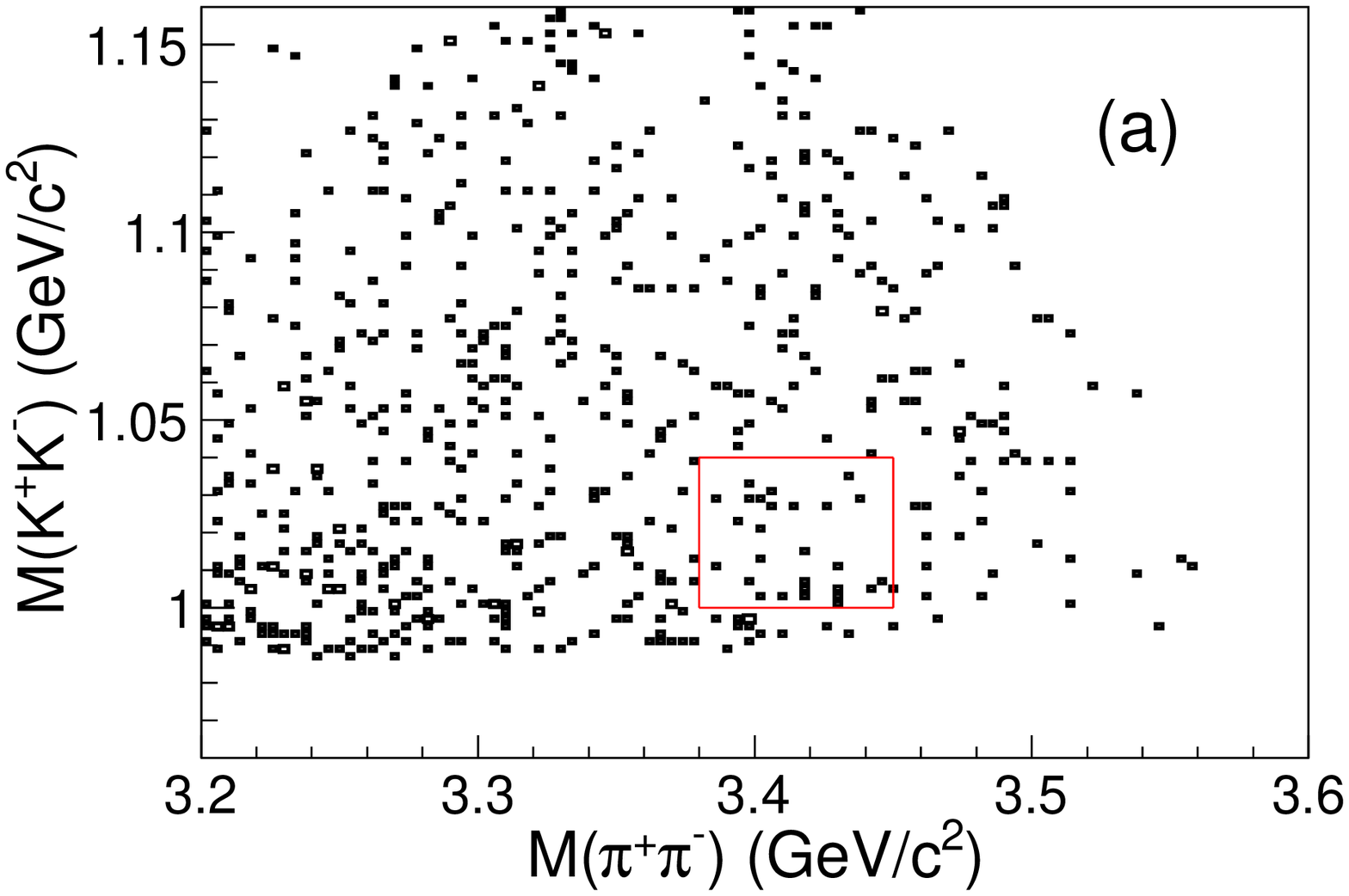}
\incfig{0.42}{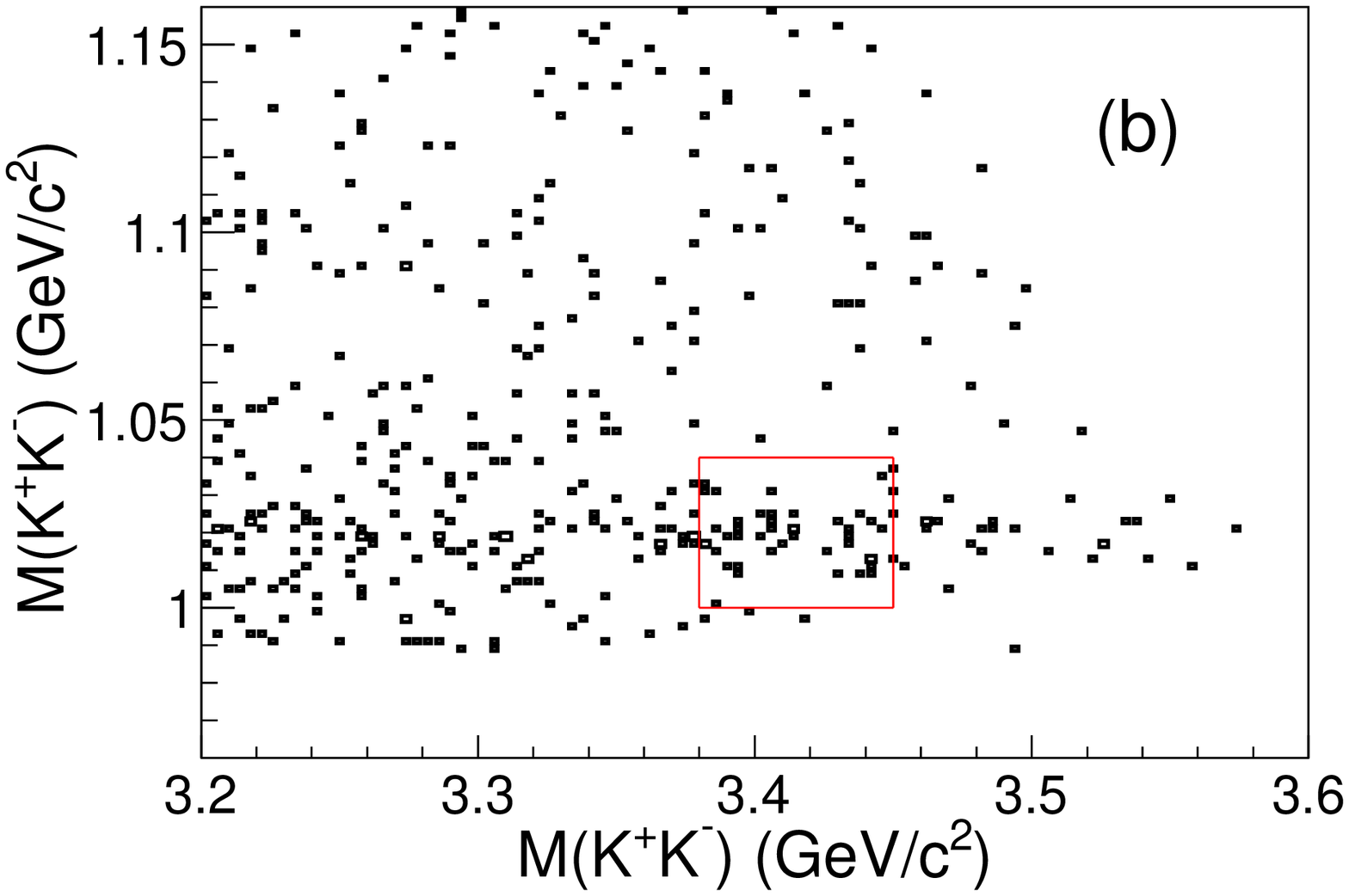}\\
\incfig{0.42}{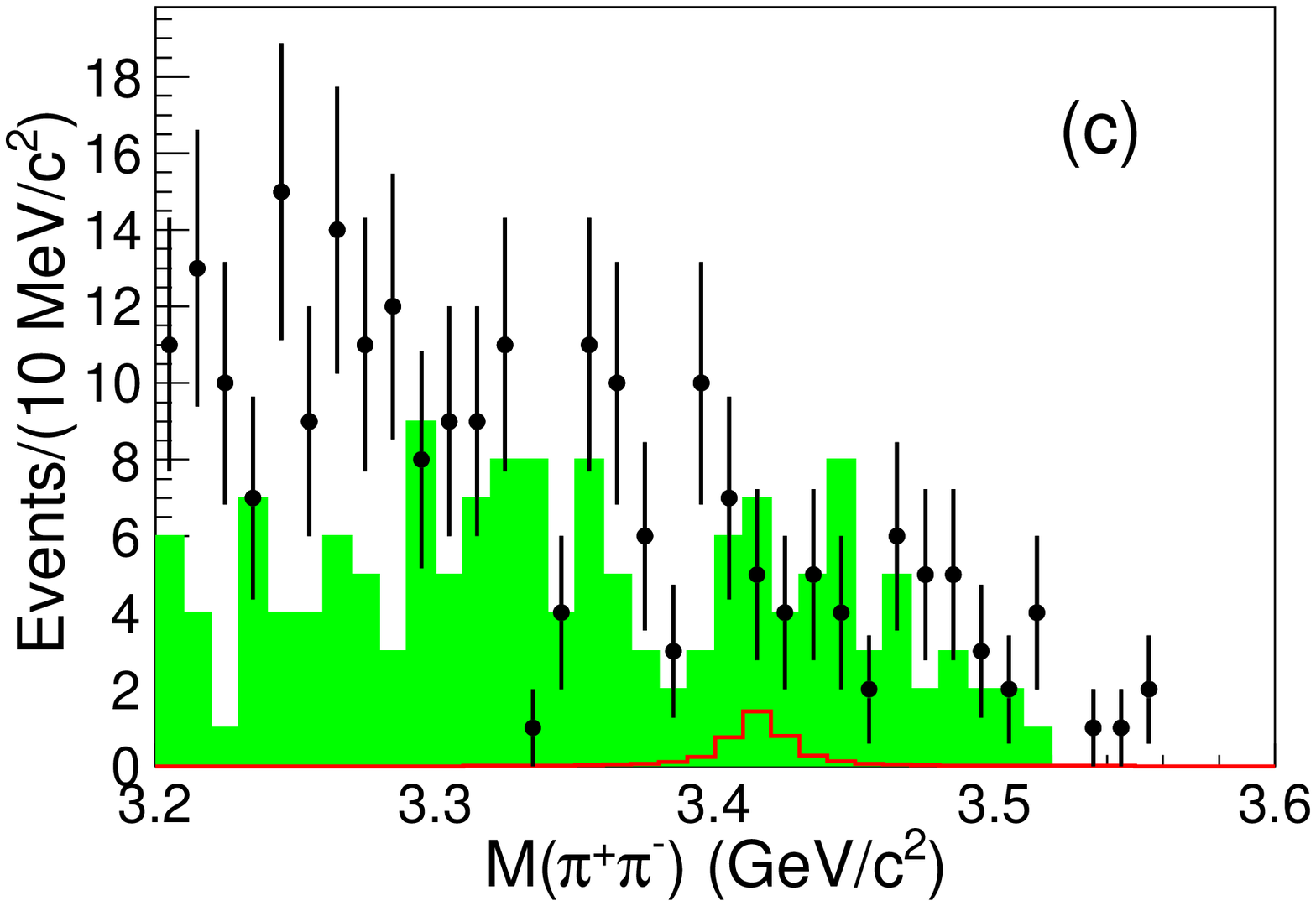}
\incfig{0.42}{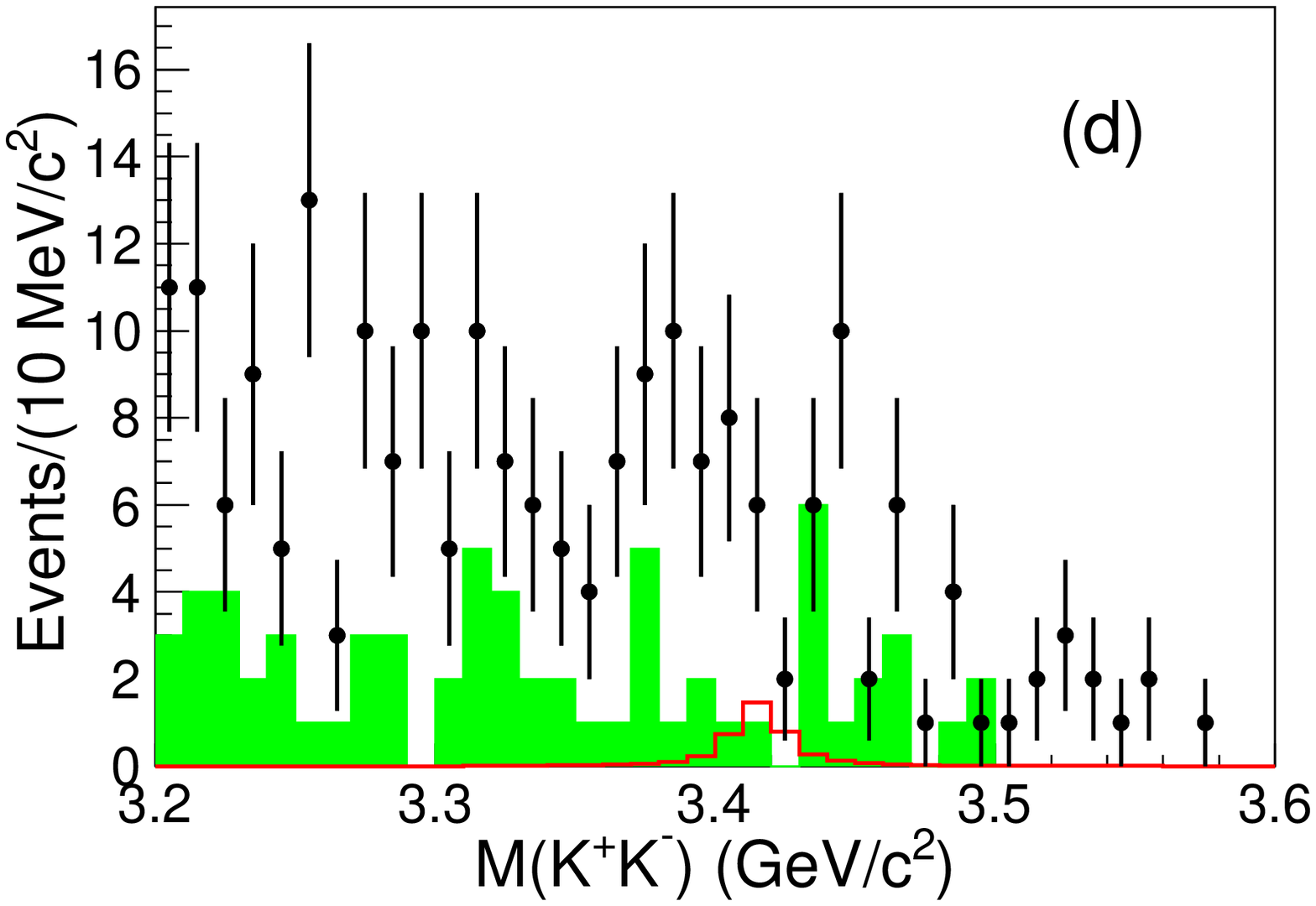}\\
\caption{(Color online) (a, b) Distributions of $M(\kk)$ for low
momentum tracks versus $M(\pp/\kk)$ for high momentum tracks, and
(c, d) the projections along $M(\pp/\kk)$ in the $\phi$ mass
window for the data sample. The red boxes represent the $\phi$ and
$\chi_{c0}$ signal regions. The dots with error bars are the data.
Histograms filled with green represent the $\phi$ sidebands, which
have been normalized to the signal region of the $\phi$. The red
histograms represent the $\chi_{c0}$ MC shape, normalized to the
upper limit of the measured cross section.}
\label{fig:scatter_kk/pp}
\end{figure*}

\subsubsection{$\chi_{c0}\to \pp\pp$}

For the $\chi_{c0}\to \pp\pp$ decay mode, the same event selection
criteria for charged tracks are applied. Four pions and only one
kaon are required to pass the PID requirement. The total charge of
the four pions is required to be zero. In order to improve the
mass resolution and suppress backgrounds, a $1$C kinematic fit is
performed with the $\ee\to K^{\pm}K^{\mp}_\text{miss}\pp\pp$
hypothesis by constraining the mass of the missing track to the
kaon mass. The $\chi^2$ of the kinematic fit is required to be
less than $20$. If there is more than one kaon, the combination of
$K^{\pm}K^{\mp}_\text{miss}\pp\pp$ with the least $\chi^{2}$ is
retained. The $\phi$ mass window is defined as above to be $0.998
\leq M(\kk) \leq 1.043$~GeV/$c^2$. Figure~\ref{fig:scatter_pppp}
shows the distribution of $M(\kk)$ versus $M(\pp\pp)$ for the data
sample and the $1$-D projections. Again, there are no obvious
$\chi_{c0}$ signals.

\subsubsection{$\chi_{c0}\to \kk\pp$}

For the $\chi_{c0}\to \kk\pp$ decay mode, we use the same criteria
to select candidate charged tracks. Two oppositely charged pions
and three kaons are required to pass the PID requirement. The
absolute value of the net charge of all kaons should not be
greater than one. A $1$C kinematic fit is performed with the
$\ee\to K^{\pm}K^{\mp}_\text{miss}\kk\pp$ hypothesis by
constraining the mass of the missing track to the kaon mass and
the $\chi^2$ of the kinematic fit is required to be less than
$20$. If there are more than three kaons, the combination of
$K^{\pm}K^{\mp}_\text{miss}\kk\pp$ with the least $\chi^{2}$ is
retained. Since the origin of the kaons from $\phi$ or $\chi_{c0}$
decays can not be determined, all combinations of $\kk$ are
considered. The $\phi$ mass window is defined as above to be
$0.998\leq M(\kk)\leq1.044$~GeV/$c^2$. The distribution of
$M(\kk)$ versus $M(\kk\pp)$ and the $1$-D projections from the
data sample are also shown in the Fig.~\ref{fig:scatter_pppp}. No
obvious $\chi_{c0}$ signals are observed.

\begin{figure*}[htbp]
\incfig{0.42}{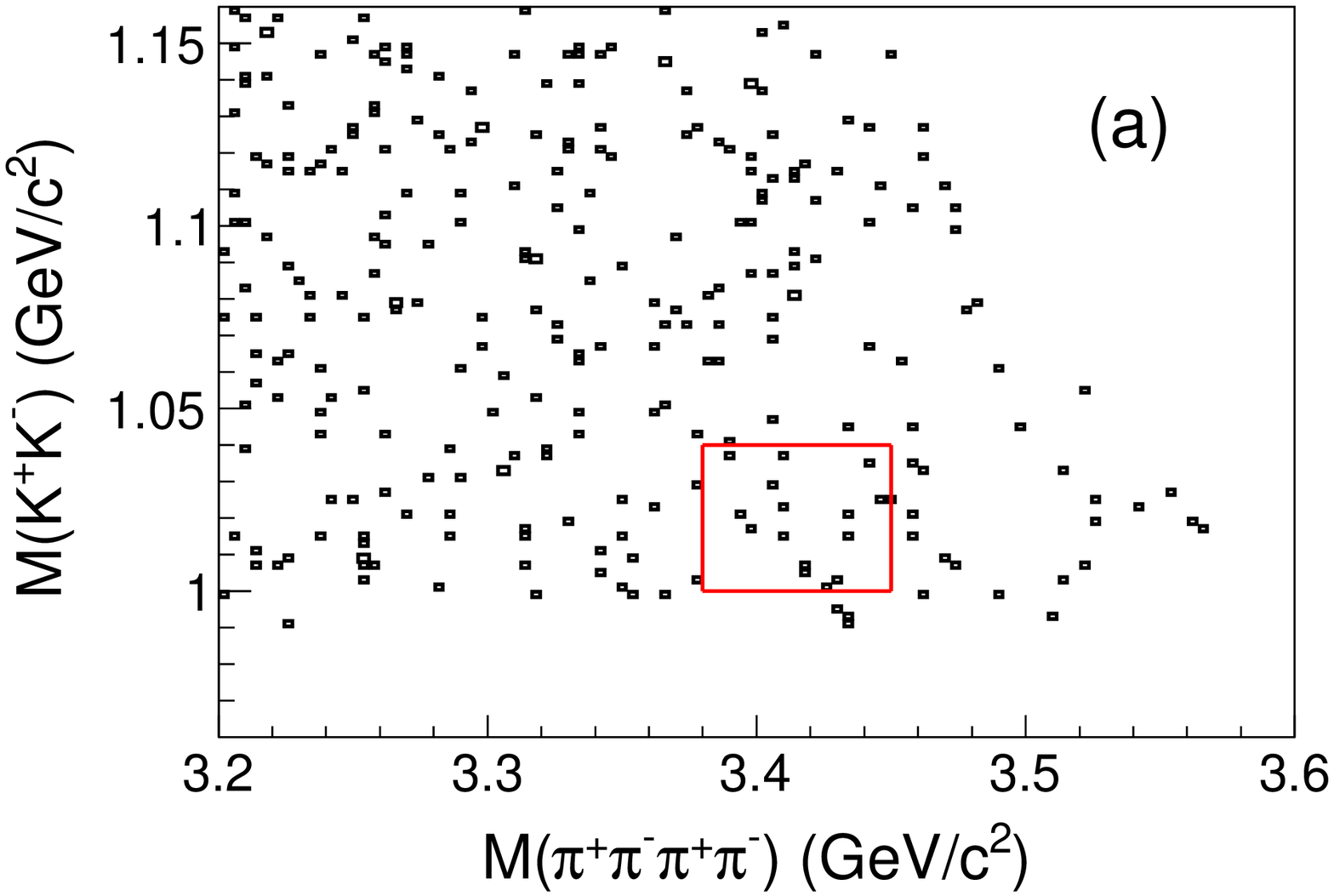}
\incfig{0.42}{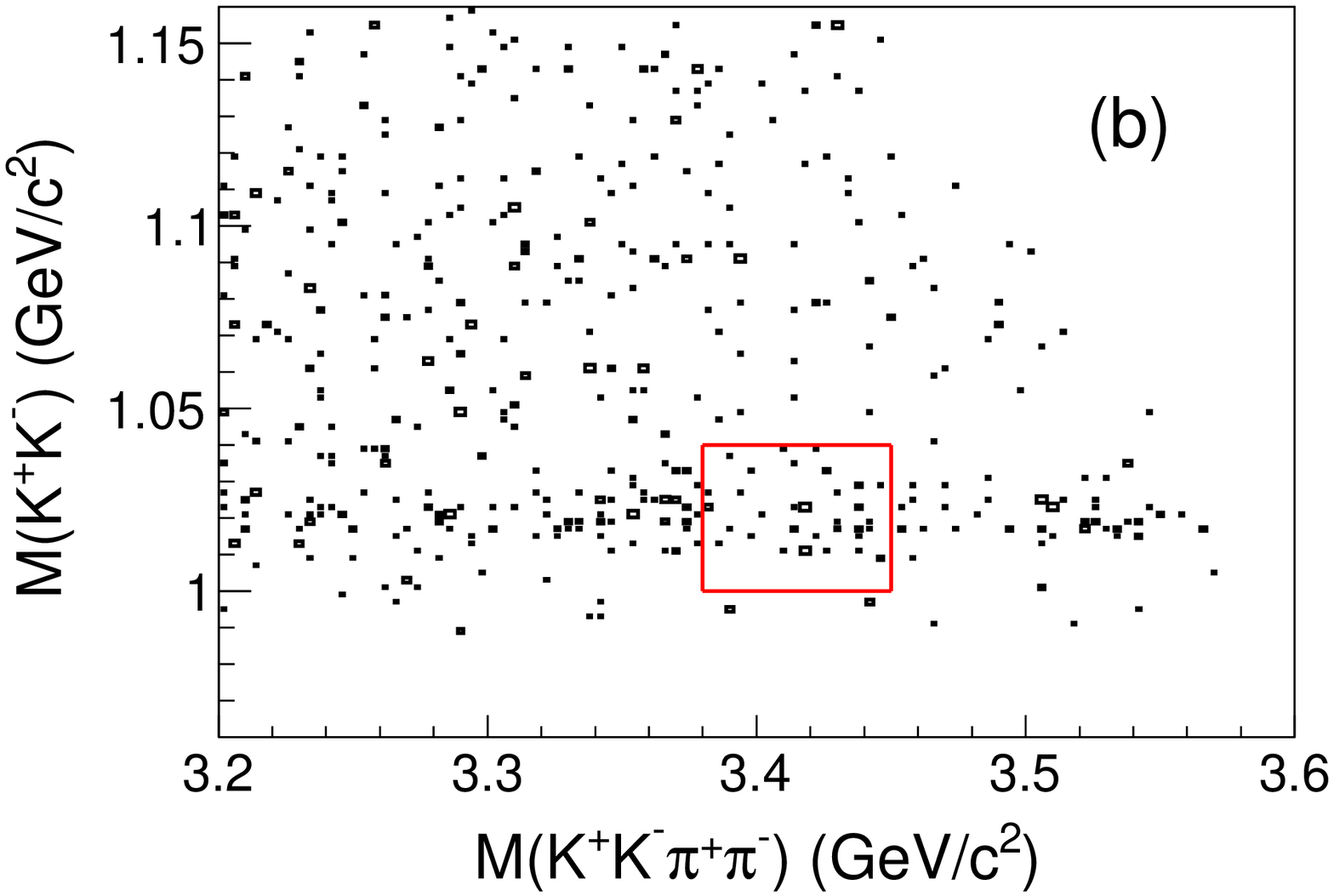}\\
\incfig{0.42}{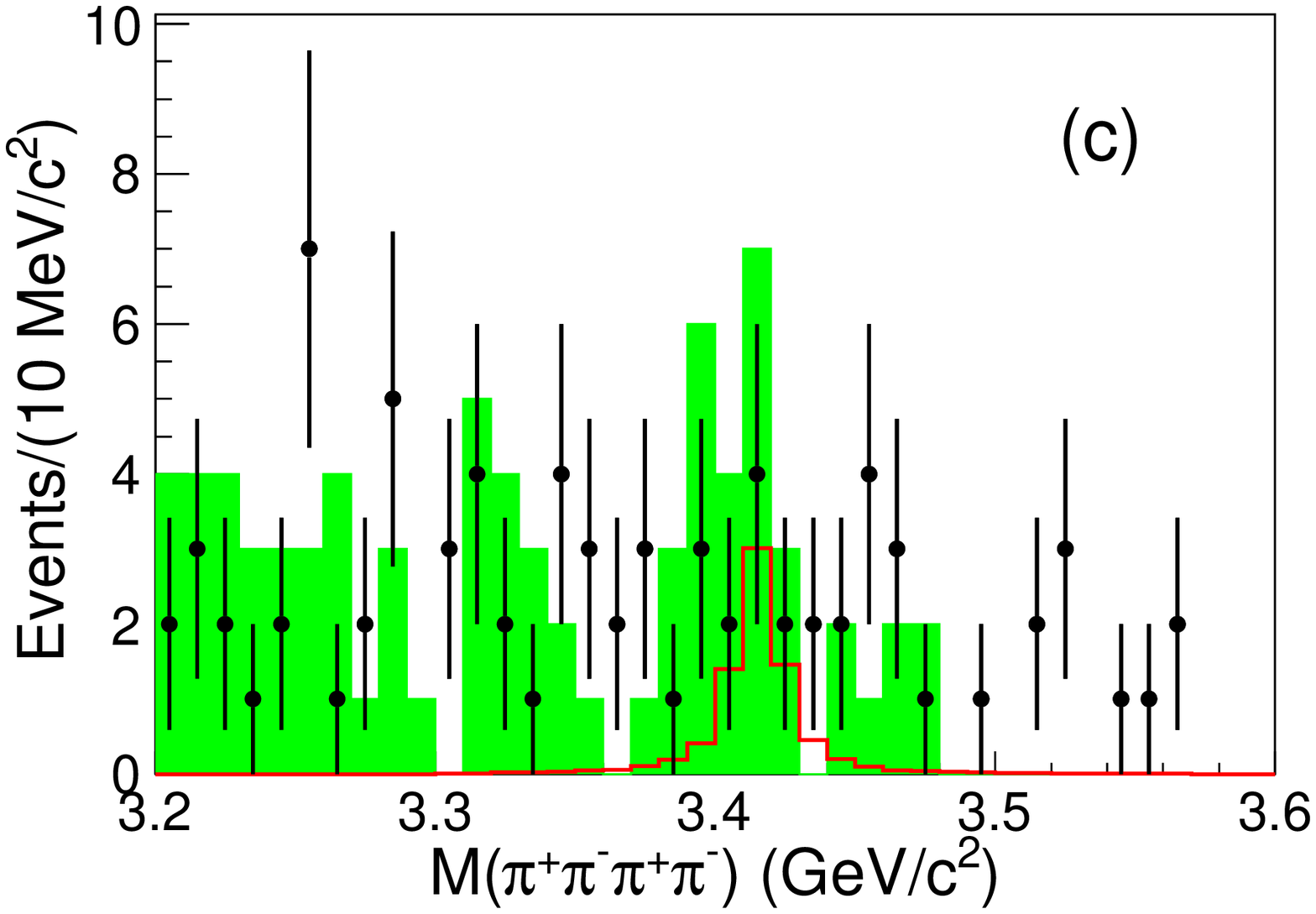}
\incfig{0.42}{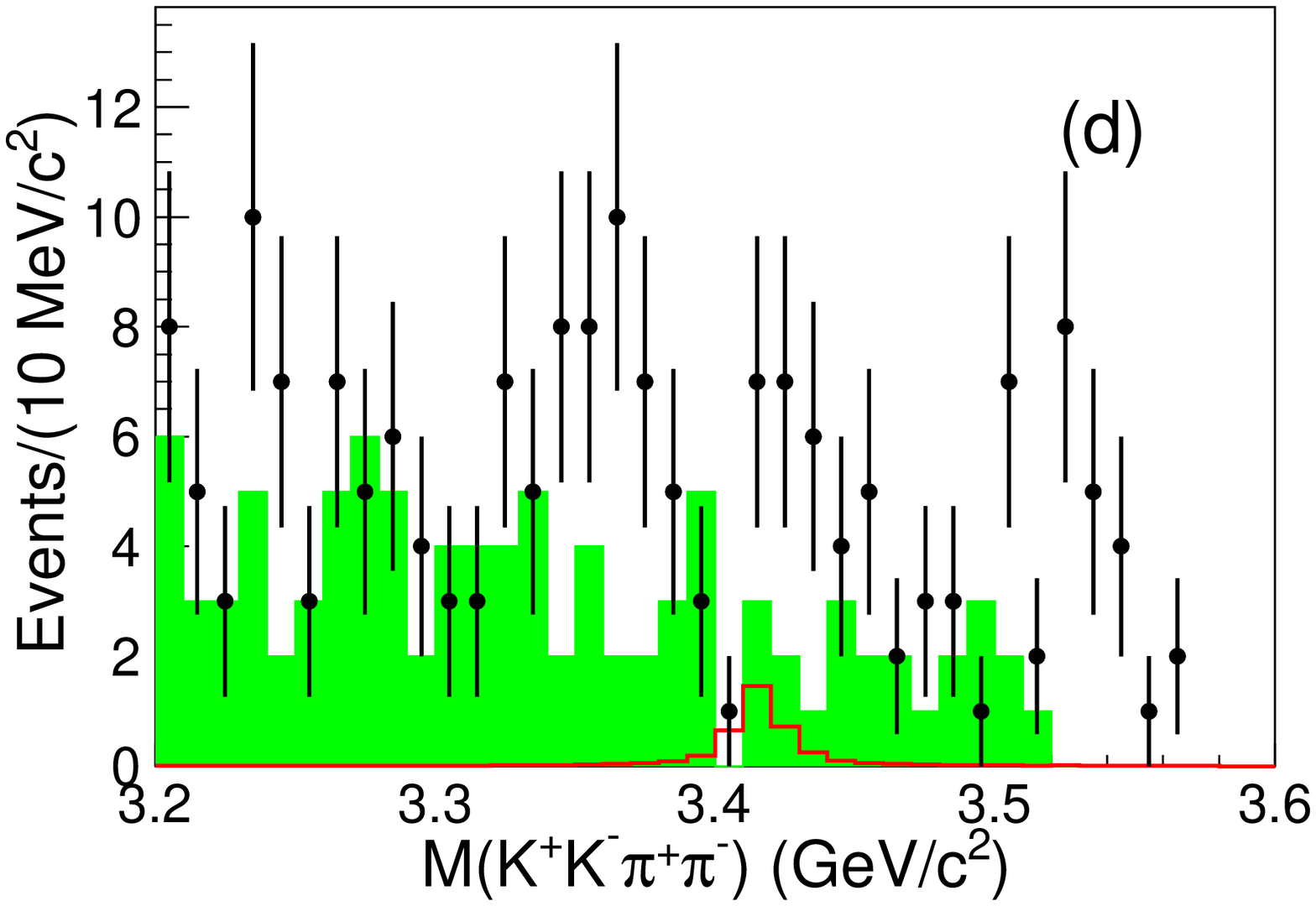}\\
\caption{(Color online) Distributions of (a) $M(\kk)$ versus
$M(\pp\pp)$, (b) $M(\kk)$ versus $M(\kk\pp)$, (c) the projection
along $M(\pp\pp)$ in $\phi$ mass window, and (d) the projection
along $M(\kk\pp)$ in $\phi$ mass window for data. The red boxes
represent the $\phi$ and $\chi_{c0}$ signal regions. The dots with
error bars are the data. The histograms filled with green
represent the $\phi$ sidebands, normalized to the signal region of
the $\phi$. The red histograms represent the $\chi_{c0}$ MC shape,
normalized to the upper limit of the measured cross section.}
\label{fig:scatter_pppp}
\end{figure*}


\subsection{Cross Section}

A simultaneous unbinned maximum likelihood fit is performed to the distributions of
$M(\pp)$, $M(\kk)$, $M(\pp\pp)$, and $M(\kk\pp)$.
The signal shape is determined from the signal MC sample, and the
background shape of each decay mode is described with a
second-order Chebychev polynomial function. The number of signal
events for each decay mode depends on its branching fraction and
efficiency. The efficiencies for $\chi_{c0}\to \pp$, $\kk$,
$\pp\pp$ and $\kk\pp$ are $62.2\%$, $58.6\%$, $29.3\%$, and
$19.7\%$, respectively. The branching fractions are obtained from
the Particle Data Group~\cite{pdg}. Since no significant
$\phi\chi_{c0}$ signal is observed, the upper limit on the Born
cross section is set at the $90\%$ confidence level (C.L.). A scan
of the likelihood with respect to the number of produced
$\phi\chi_{c0}$ events is obtained, and the upper limit on $n^{\rm
prod}$ at the $90\%$ C.L.\ is determined according to
$\int^{n^{\rm prod}}_{0} L(x)dx/\int^{\infty}_{0}L(x)dx = 0.9$.
Since the branching fractions and efficiencies of the four decay
modes have been considered in the fit, the upper limit on the Born
cross section is calculated with
\begin{equation}
\label{equ:up1} \sigma^{\rm B}=\frac{n^{\rm
prod}}{\mathcal{L}_{\rm int} (1+\delta)(1+\delta^{\rm vac})},
\end{equation}
where $(1+\delta)=0.74$~\cite{radiators} and  $(1+\delta^{\rm
vac})=1.055$~\cite{vacuum} obtained with the same method as for
$\ee\to \phi\chi_{c1,2}$. The upper limit on $\sigma^{\rm B}$ is
obtained by replacing $n^{\rm prod}$ with that on $n^{\rm prod}$.
To take the systematic uncertainty into account, the likelihood
distribution is convolved with a Gaussian function with a mean
value of $0$ and a standard deviation of $n^{\rm
prod}\cdot\Delta$, where $n^{\rm prod}$ is the number of produced
$\ee\to \phi\chi_{c0}$ events and $\Delta$ is the relative
systematic uncertainty described in next section. The upper limit
on the production of $\ee\to \phi\chi_{c0}$ at the $90\%$ C.L.\ is
estimated to be $5.4$~pb.


\section{\boldmath $\ee\to \gamma X(4140)$}

For $\ee\to \gamma X(4140)$, we search for $X(4140)$ meson decays
to $\phi\jpsi$, with $\jpsi$ decaying to $\LL$, and $\phi$
decaying to $\kk$. Since the final state of $\ee\to \gamma
X(4140)$ is the same as that for $\ee\to \phi\chi_{c1,2}$, we
apply the same event selection criteria and requirements.  The
resulting distributions $M(\phi\jpsi)$ and $M(\gamma\jpsi)$ in
the $\phi$ and $\jpsi$ mass windows are shown in
Fig.~\ref{fig:m4140}. An unbinned maximum likelihood fit is
performed to the distribution of $M(\gamma\jpsi)$. The signal
shape is determined from the signal MC sample and the background
shapes are described with those from MC simulations for $\ee\to
\phi\chi_{c1}$ and $\phi\chi_{c2}$. Since there is no obvious
$X(4140)$ signal, the upper limit on the Born cross section at the
$90\%$ C.L.\ is determined. The upper limit on the number of
signal events is obtained with the same method as for $\ee\to
\phi\chi_{c0}$. The upper limit on the Born cross section is
calculated using Eq.~(\ref{equ:born}),  where
$(1+\delta)=0.75$~\cite{radiators} and $(1+\delta^{\rm
vac})=1.055$~\cite{vacuum} obtained with the method described
above. The upper limit on the production of the Born cross section
and branching fraction $\sigma[\ee\to \gamma
X(4140)]\cdot\mathcal{B}(X(4140)\to \phi\jpsi)$ at the $90\%$
C.L.\ is estimated to be $1.2$~pb. The distribution of
$M(\phi\jpsi)$ is also fitted, but a higher upper limit is
obtained. Toy MC samples with the two methods are generated and
studied, and we obtain a better sensitivity when applying the fit
to $M(\gamma\jpsi)$.

\begin{figure*}[htbp]
\incfig{0.42}{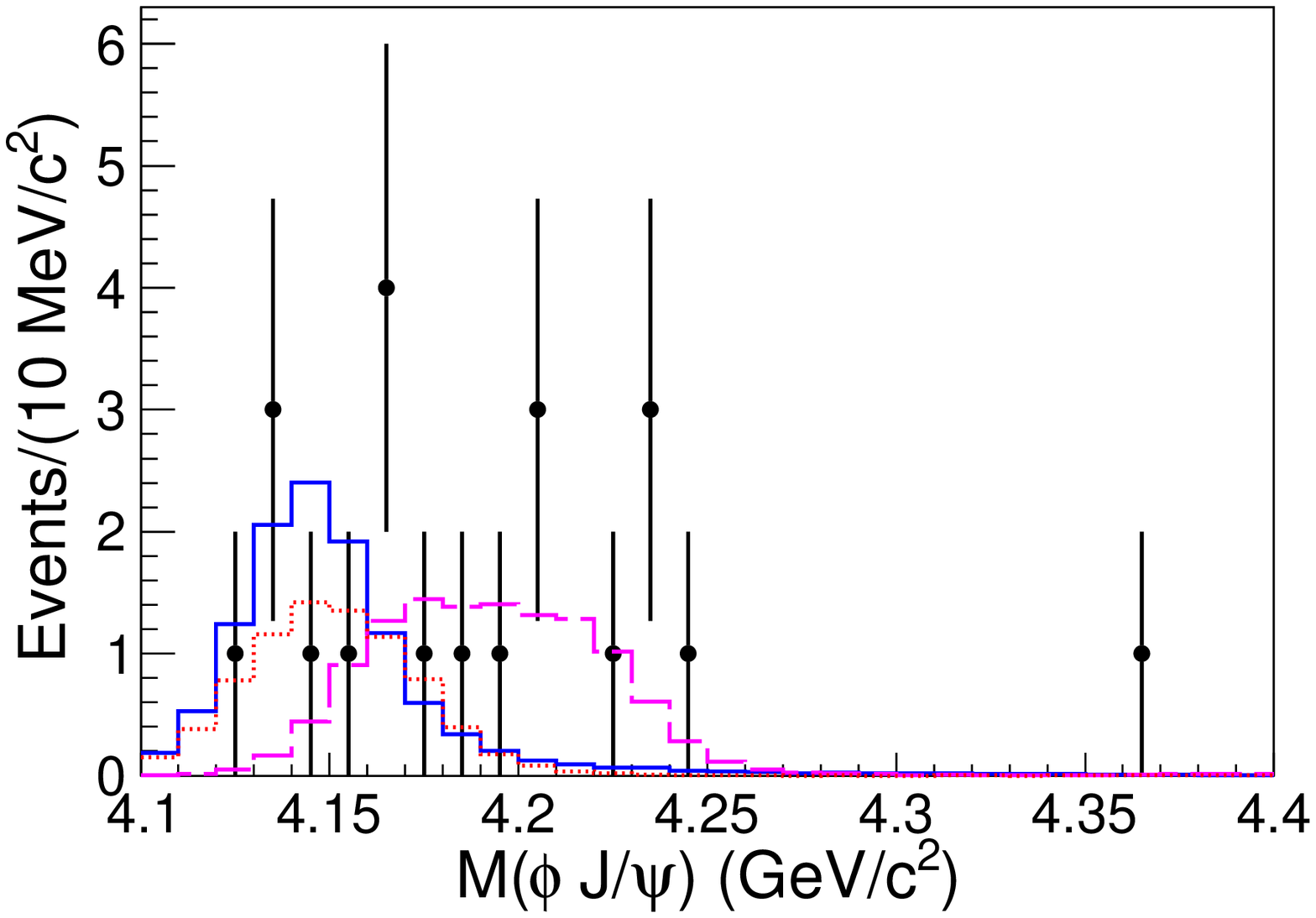}
\incfig{0.42}{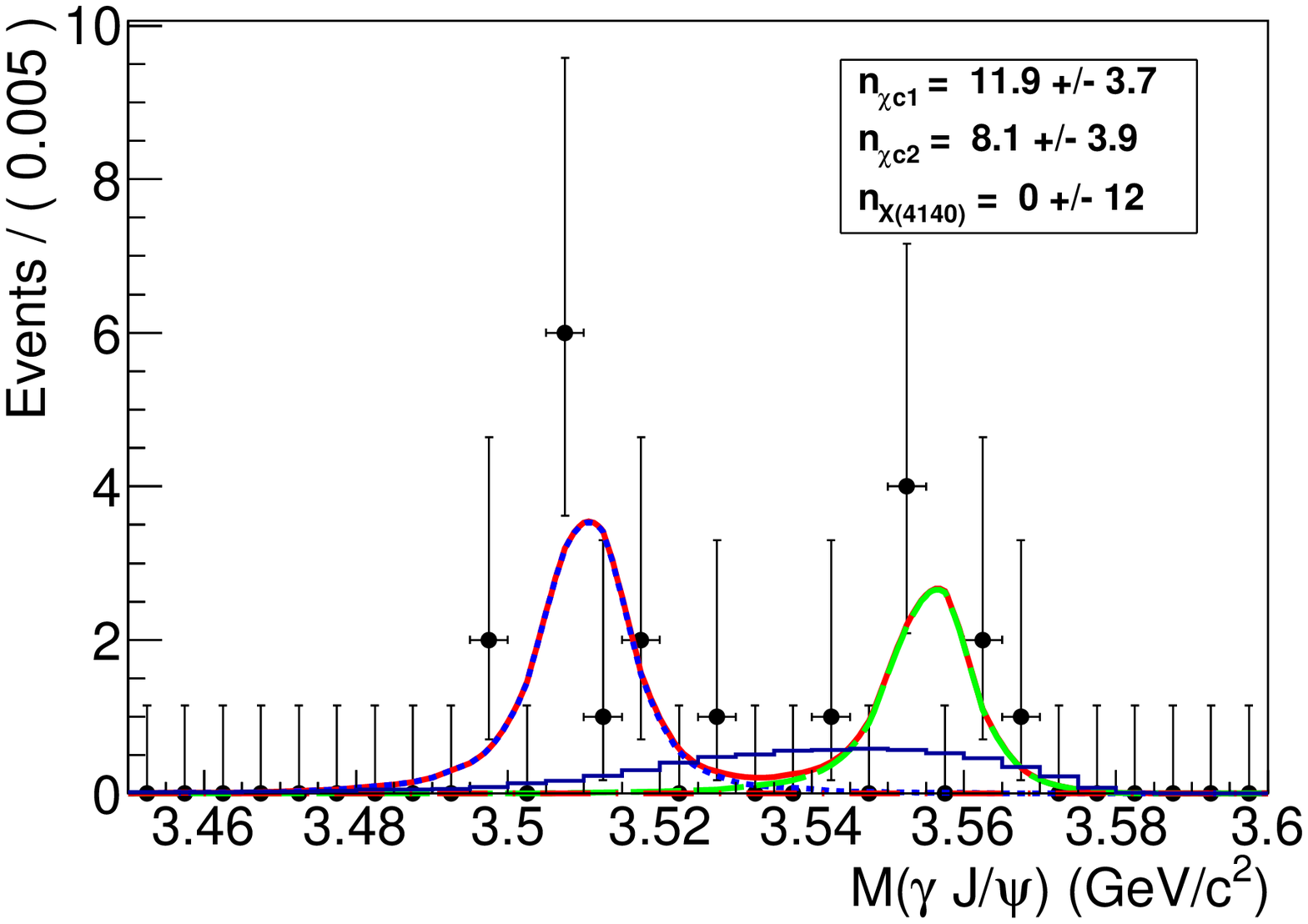}\\
\caption{(Color online) (Left) Distribution of $M(\phi\jpsi)$
in the $\phi$ and $\jpsi$ mass windows for data. The
dots with error bars are the data and the blue solid histogram
represents the MC shape from $M(\gamma X(4140))$, normalized to the
upper limit of the Born cross section. The magenta long-dashed and
red dotted histograms represent the MC shapes from $M(\phi\chi_{c1})$
and $M(\phi\chi_{c2})$, respectively, normalized to the measured Born
cross sections. (Right) Fit to the distribution of $M(\gamma\jpsi)$.
Dots with error bars are data. Red solid line is
the fit curve. Blue dashed and green long-dashed lines represent
$\chi_{c1}$ and $\chi_{c2}$ backgrounds, respectively. Red
dash-dotted line represents $X(4140)$ signal. The blue histogram
represents the $X(4140)$ signal shape from MC simulation with
arbitrary normalization.} \label{fig:m4140}
\end{figure*}


\section{Systematic Uncertainty}

The systematic uncertainties on the cross section measurements for
$\ee\to \phi\chi_{c0,1,2}$ and $\ee\to \gamma X(4140)$ come mainly from
the integrated luminosity, the tracking and photon reconstruction, the PID,
the kinematic fit, the signal and background shapes, the fit range, the branching
fraction and the radiative correction. The systematic uncertainties are
summarized in Table~\ref{tab:syserror} and explained below.

\begin{table*}[hbp]
\caption{\small{The relative systematic uncertainties of Born
cross sections (\%) for $\ee\to \phi\chi_{c0,1,2}$ and $\ee\to
\gamma X(4140)$ at $\sqrt{s}=4.600$~GeV.} An ellipsis ($\cdots$)
means that the uncertainty is negligible.} \label{tab:syserror}
\begin{tabular}{ccccc}
  \hline
  \hline
  \multirow{1}{0.15\textwidth}{\centering{Source}} & \multirow{1}{0.13\textwidth}{\centering{$\phi\chi_{c0}$}} & \multirow{1}{0.13\textwidth}{\centering{$\phi\chi_{c1}$}} & \multirow{1}{0.13\textwidth}{\centering{$\phi\chi_{c2}$}} & \multirow{1}{0.13\textwidth}{\centering{$\gamma X(4140)$}}\\
  \hline
  Luminosity            &    $1.0$            &      $1.0$           &    $1.0$              &         $1.0$\\
  Tracking              &    $4.2$            &      $3.0$           &    $3.0$              &         $3.0$\\
  Photon                &    $\cdots$         &      $1.0$           &    $1.0$              &         $1.0$\\
  PID                   &    $3.4$            &      $1.0$           &    $1.0$              &         $1.0$\\
  Kinematic fit         &    $1.6$            &      $1.5$           &    $1.0$              &         $2.4$\\
  Branching fraction    &    $5.7$            &      $3.8$           &    $3.9$              &         $1.2$\\
  Radiative correction  &    $5.2$            &      $2.1$           &    $2.2$              &         $7.3$\\
  Angular distribution  &    $3.7$            &      $4.5$           &    $4.3$              &         $13.8$\\
  Signal shape          &    $3.4$            &      $\cdots$        &    $\cdots$           &         $11.0$\\
  Background shape      &    $5.2$            &      $\cdots$        &    $\cdots$           &         $\cdots$\\
  Fitting range         &    $1.0$            &      $\cdots$        &    $\cdots$           &         $1.7$\\
  \hline
  Sum                   &    $12.1$           &      $7.3$           &    $7.2$              &         $19.7$\\
  \hline
  \hline
\end{tabular}
\end{table*}

The systematic uncertainty due to the detection efficiency includes
uncertainties from track reconstruction, PID efficiency, photon
reconstruction, the kinematic fit, angular distributions and the radiative
correction. The uncertainty from track reconstruction for each charged
track is taken as $1.0\%$~\cite{tracks}. In the process $\ee\to \phi\chi_{c0}$, the total
systematic uncertainty from tracking reconstruction is obtained by taking
into account the weights of the efficiencies and branching fractions of the
four $\chi_{c0}$ decay modes. The total systematic uncertainty due to PID
efficiency is obtained with the same method, where the PID uncertainty for
each charged track is taken as $1.0\%$~\cite{tracks}. The systematic uncertainty from
photon reconstruction is determined to be $1.0\%$ for each photon by studying
the control sample of $\jpsi\to \rho^0\pi^0$ decays~\cite{photon}.

Since it is difficult to find an appropriate control sample to
estimate the systematic uncertainty related to the kinematic fit
and the vertex fit, we correct the charged track helix parameters
of the MC simulated events~\cite{kinematic} to obtain a better
match with the data sample. The difference between the efficiency
with and without the correction is taken as the uncertainty
associated with the kinematic fit. The MC sample with the track
helix parameter correction applied is used in the nominal
analysis.

In order to estimate the uncertainty from the angular
distributions of the $\phi$ meson and the radiative photon, we
change the decay dynamics from phase space to $1+\cos^2\theta$ or
$1-\cos^2\theta$ to generate new signal MC samples. For $\ee\to
\gamma X(4140)$, $\theta$ is the polar angle of the radiative
photon in the $\ee$ rest frame with the $z$ axis pointing in the
direction of the electron beam, while for $\ee\to
\phi\chi_{c0,1,2}$, $\theta$ is the polar angle of the $\phi$
meson. The maximum difference in efficiency is taken as the
systematic uncertainty.

The line shape used in the MC simulation will affect both the
radiative correction factor and the efficiency. In the nominal MC
simulation, we assume that the processes $\ee\to
\phi\chi_{c0,1,2}$ and $\ee\to \gamma X(4140)$ follow the line
shape of the $X(4660)$~\cite{4660} modified by a phase space
factor. We change the line shape to
$\frac{4\pi\alpha^2}{3s}\Phi(\sqrt{s})$ and the resultant
difference of $(1+\delta)\cdot\epsilon$ is taken as the systematic
uncertainty due to the radiative correction factor.

The luminosity is measured using large-angle Bhabha events with an
uncertainty of less than $1.0\%$~\cite{luminosity}. The branching
fractions for $\phi\to \kk$, $\chi_{c1,2}\to \gamma\jpsi$,
$\jpsi\to \LL$ and $\chi_{c0}\to \pp$, $\kk$, $\pp\pp$, $\kk\pp$
are taken from the Particle Data Group~\cite{pdg}. The
uncertainties of the branching fractions are taken as the
associated systematic uncertainties. For the $\phi$ and $\jpsi$
mass windows, very loose criteria are used, hence the difference
in efficiency between MC simulation and data sample is negligible.

The yields of signal $\ee\to \phi\chi_{c0}$ and $\ee\to \gamma
X(4140)$ are determined from the fit, and the yields of signal of
$\ee\to \phi\chi_{c1,2}$ is obtained by simply counting events.
Only the systematic uncertainty associated with the fit is
considered. The systematic uncertainty on the fit procedure
comprises those due to the signal shape, background shape and fit
range. For $\ee\to \phi\chi_{c0}$, we generate alternative signal
MC samples by varying the mass and width of the $\chi_{c0}$ by one
standard deviation and take the maximum difference with respect to
the nominal values as the systematic uncertainty due to the signal
shape. The systematic uncertainty caused by the background shape
is obtained by changing the background shape from a second-order
polynomial function to a third-order polynomial function. The
nominal fit range is taken to be $[3.18, 3.58]$~GeV/$c^2$. We vary
the limit of the fit range by $\pm 0.05$~GeV/$c^2$ and take the
difference as the associated systematic uncertainty. For $\ee\to
\gamma X(4140)$, we generate a signal MC sample by varying the
mass and width of the $X(4140)$ with one standard deviation and
take the maximum difference as the systematic uncertainty due to
the signal shape. The nominal fit range is taken to be $[3.45,
3.60]$~GeV/$c^2$. We vary the limit of the fit range by
$\pm0.01$~GeV/$c^2$ and take the resultant difference as the
associated systematic uncertainty.

The total systematic uncertainties are obtained by adding the
individual uncertainties in quadrature, assuming that all sources
are independent. For $\ee\to \phi\chi_{c0,1,2}$ and $\ee\to \gamma
X(4140)$, the total systematic uncertainties are $12.1\%$,
$7.3\%$, $7.2\%$ and $19.7\%$, respectively.


\section{RESULTS AND DISCUSSION}

In summary, the processes $\ee\to \phi\chi_{c1}$ and
$\phi\chi_{c2}$ are observed for the first time at a c.m.\ energy
of $\sqrt{s}=4.600$~GeV by using a data sample corresponding to an
integrated luminosity of $567$~$\rm pb^{-1}$ collected with the
BESIII detector. The corresponding Born cross sections are
measured to be $(4.2^{+1.7}_{-1.0}\pm 0.3)$~pb and
$(6.7^{+3.4}_{-1.7}\pm 0.5)$~pb, respectively. No obvious signals
are observed for $\ee\to \phi\chi_{c0}$ and $\ee\to \gamma
X(4140)$ and the upper limits on the Born cross sections at the
$90\%$ C.L.\ are set to be $5.4$~pb and $1.2$~pb, respectively.

Since only one data set at or near $\sqrt{s}=4.600$~GeV is
available to study these modes at BESIII, it is not possible to
measure the line shape for their production. The cross sections of
other decay modes at this energy point, such as $\ee\to \pp\jpsi$,
$\ee\to \omega\chi_{c0,1,2}$, are all at the level of a few pb. As
$\ee\to \phi\chi_{c1,2}$ signals have been observed, it will be
interesting to measure the line shape between the threshold to
4.600~GeV or even higher.

The upper limit of the Born cross section for $\ee\to\gamma
X(4140)$ at $4.600$~GeV is higher than those measured at $4.230$,
$4.260$, and $4.360$~GeV, due to the non-trivial backgrounds from
$\chi_{c1,2}$. Measurements based on data samples with larger
statistics at more energy points will be helpful to clarify the
nature of these decay processes in this energy region.

\acknowledgments

The BESIII collaboration thanks the staff of BEPCII and the IHEP
computing center for their strong support. This work is supported
in part by National Key Basic Research Program of China under Contract
No. 2015CB856700; National Natural Science Foundation of China (NSFC)
under Contracts Nos. 11235011, 11335008, 11425524, 11625523, 11635010;
the Chinese Academy of Sciences (CAS) Large-Scale Scientific Facility
Program; the CAS Center for Excellence in Particle Physics (CCEPP);
Joint Large-Scale Scientific Facility Funds of the NSFC and CAS under
Contracts Nos. U1232106, U1332201, U1532257, U1532258; CAS Key Research
Program of Frontier Sciences under Contracts Nos. QYZDJ-SSW-SLH003,
QYZDJ-SSW-SLH040; 100 Talents Program of CAS; National 1000 Talents
Program of China; INPAC and Shanghai Key Laboratory for Particle Physics
and Cosmology; German Research Foundation DFG under Contracts Nos.
Collaborative Research Center CRC 1044, FOR 2359; Istituto Nazionale
di Fisica Nucleare, Italy; Koninklijke Nederlandse Akademie van
Wetenschappen (KNAW) under Contract No. 530-4CDP03; Ministry of
Development of Turkey under Contract No. DPT2006K-120470; National
Science and Technology fund; The Swedish Research Council; U. S.
Department of Energy under Contracts Nos. DE-FG02-05ER41374,
DE-SC-0010118, DE-SC-0010504, DE-SC-0012069; University of
Groningen (RuG) and the Helmholtzzentrum fuer Schwerionenforschung
GmbH (GSI), Darmstadt; WCU Program of National Research Foundation
of Korea under Contract No. R32-2008-000-10155-0; Shandong Natural
Science Funds for Distinguished Young Scholar under Contract No. JQ201402

\input{bibliography_prd}
\end{document}

%% file: authors_feb2017.tex
\author{
M.~Ablikim$^{1}$, M.~N.~Achasov$^{9,d}$, S. ~Ahmed$^{14}$, M.~Albrecht$^{4}$, A.~Amoroso$^{53A,53C}$, F.~F.~An$^{1}$, Q.~An$^{50,40}$, J.~Z.~Bai$^{1}$, Y.~Bai$^{39}$, O.~Bakina$^{24}$
% ...
\\
\vspace{0.2cm}
(BESIII Collaboration)\\
\vspace{0.2cm} {\it
$^{1}$ Institute of High Energy Physics, Beijing 100049, People's Republic of China\\
% ...
$^{4}$ Bochum Ruhr-University, D-44780 Bochum, Germany\\
% ...
$^{9}$ G.I. Budker Institute of Nuclear Physics SB RAS (BINP), Novosibirsk 630090, Russia\\
% ...
$^{14}$ Helmholtz Institute Mainz, Johann-Joachim-Becher-Weg 45, D-55099 Mainz, Germany\\
% ...
$^{24}$ Joint Institute for Nuclear Research, 141980 Dubna, Moscow region, Russia\\
% ...
$^{39}$ Southeast University, Nanjing 211100, People's Republic of China\\
$^{40}$ State Key Laboratory of Particle Detection and Electronics, Beijing 100049, Hefei 230026, People's Republic of China\\
% ...
$^{50}$ University of Science and Technology of China, Hefei 230026, People's Republic of China\\
% ...
$^{53}$ (A)University of Turin, I-10125, Turin, Italy; (B)University of Eastern Piedmont, I-15121, Alessandria, Italy; (C)INFN, I-10125, Turin, Italy\\
% ...
$^{d}$ Also at the Novosibirsk State University, Novosibirsk, 630090, Russia\\
% ...
}
}